\documentclass{PoS}

\title{Observing future Solar Flare Neutrino in Hyper-KamioKande in Japan, Korea and in IceCube}

\ShortTitle{Solar Flare in Megaton HK in Japan, Korea, IceCube}

\author{\speaker{Daniele Fargion}\\
       Physics Department \& INFN Rome1, Rome University 1, P.le A. Moro 2, 00185, Rome, Italy\\
       MIFP, Via Appia Nuova 31, 00040 Marino (Rome), Italy\\
       E-mail: \email{daniele.fargion@roma1.infn.it}}

\author{Pietro Oliva\\
      Niccol\`o Cusano University, Via Don Carlo Gnocchi 3, 00166 Rome, Italy\\
      MIFP, Via Appia Nuova 31, 00040 Marino (Rome), Italy\\
      Department of Sciences, University Roma Tre, Via Vasca Navale 84, 00146 Rome, Italy\\
        E-mail: \email{pietro.oliva@unicusano.it}}

\FullConference{35th International Cosmic Ray Conference\\
	 10-20 July, 2017\\
	 Bexco, Busan, Korea}

\usepackage{journals, amsmath}
\usepackage{lipsum}

\abstract{
The largest Solar flare have been recorded in gamma flash and hard spectra up to tens GeV energy. The present building and upgrade of Hyper-Kamiokande (HK) and IceCube (as well as Deep Core, PINGU) neutrino detectors do offer a novel way to largest trace of solar Flare: their sudden anti-neutrino (or neutrino) flare made by proton scattering and pion decays via Delta resonance production. These signals might be observable at largest flare by HK via soft spectra up to tens-hundred MeV energy and by IceCube-PINGU at higher, GeV up to hundred GeV, energies. We show the expected rate of signals for the most powerful solar flare occurred in recent decades.
}

\begin{document}

\begin{figure}[!b]
\begin{center}
\includegraphics[scale=0.56]{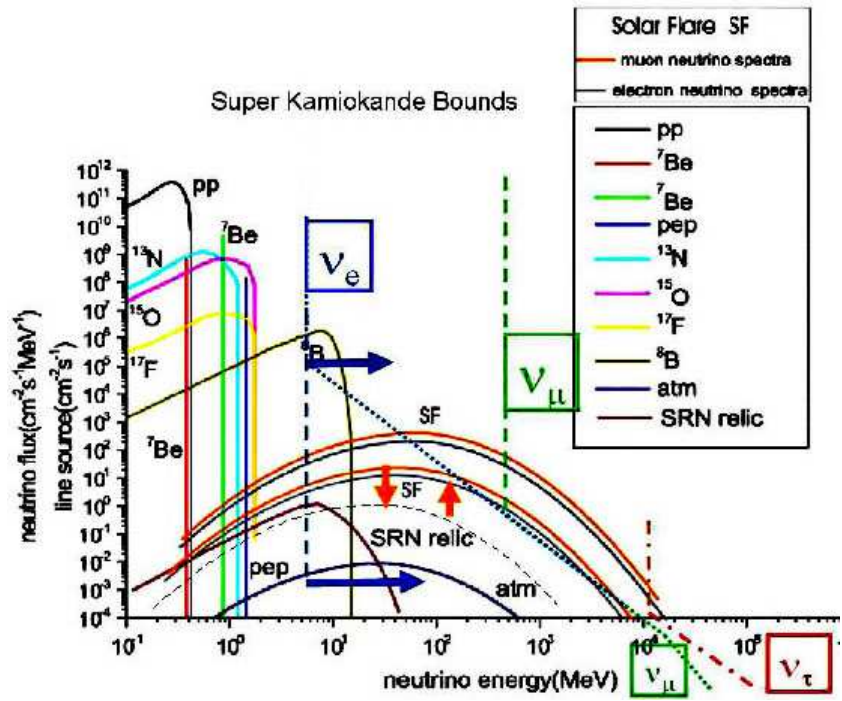}
\caption{The whole electron neutrino spectra with the main solar neutrino component. The different flavor energy threshold is also marked by arrows.} \label{Fig1}
\end{center}
\end{figure}
\section{Introduction}

During the largest solar flare, of a few minutes duration, the particle flux escaping the corona eruption, and hitting later on the Earth, is of 3-4 order of magnitude above the common atmospheric CR  background. If the flare particle interactions on the Sun corona is taking place as efficiently as they do in our terrestrial atmosphere, than their earliest proton-proton scattering, their pion  secondaries  and their muons decays ejected in solar corona are leading to a prompt neutrino fluency on Earth comparable to one day integral terrestrial atmospheric
neutrino activity (upper Bound).
One therefore may expect a prompt increase of the neutrino signals of the order of one day integral events (of atmospheric neutrinos)
made by such rarest solar flare neutrinos. Therefore there may exist a prompt solar flare neutrino astronomy.
\cite{Fargion_2003, Fargion_2004, Fargion_2006}.
In present neutrino detectors the signal is just on the edge, but as long as the authors know, it has been never revealed
\cite{Fargion_2004, Fargion_2006}. Sun density at the flare corona might be diluted and pion production maybe consequently suppressed by a factor 0.1-0.05 with respect to terrestrial atmosphere, leading to a signal at few percent the expected one under the
above considerations. This may be the reason for the SK null detection. Indeed the low Gamma signals recently reported
\cite{Grechnev 2008} confirm this suppressed signal, but just at the detection edge. Unfortunately the neutrino
signal at hundred MeV energies is rare while the one at ten MeV or below is polluted by Solar Hep neutrinos. The expected
signal is dominated by 10-30~MeV neutrinos, that might be greatly improved by anti-neutrino component via
Gadolinium presence in next SK detectors \cite{Fargion_2006}. Our earliest (2006) upper limit estimate for October - November 2003
solar flares and the recent January 20th 2005 exceptional flare were leading to signal near unity for Super-
Kamiokande and to several events well above unity for any future Megaton detector.

\begin{figure}[!t]
\begin{center}
\includegraphics[scale=0.27]{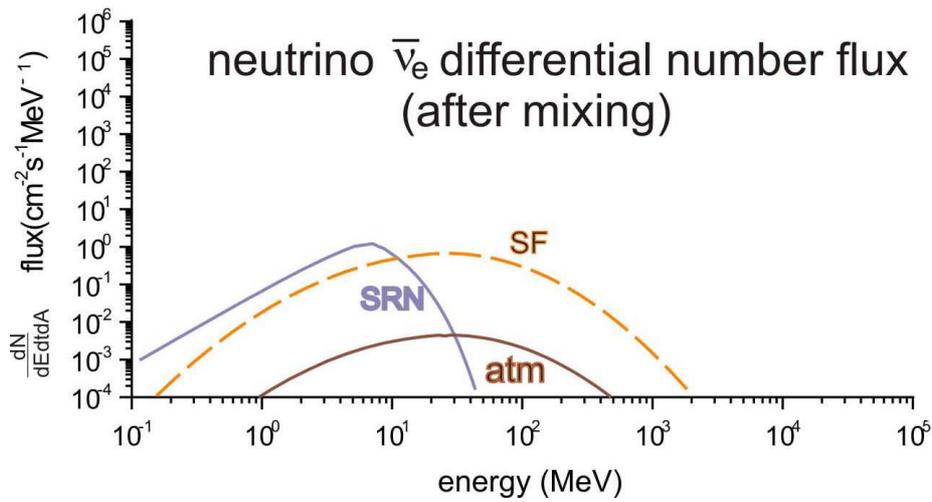}
\caption{The whole electron anti neutrino spectra without the main solar neutrino noise. The different flavor energy threshold is also marked by the arrows.} \label{Fig1b}
\end{center}
\end{figure}
\begin{figure}[!t]
\begin{center}
\includegraphics[width=12cm]{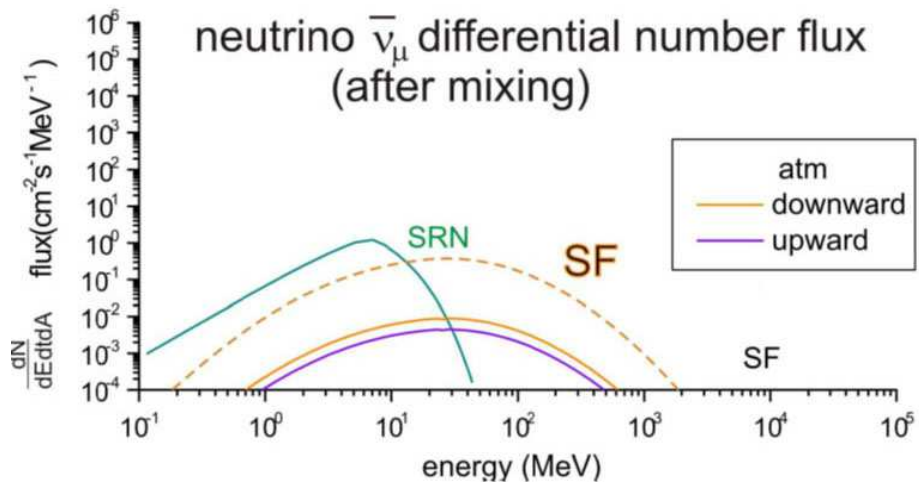}
\caption{The whole muon anti.neutrino spectra with the main solar neutrino component almost absent. The different flavor energy threshold is also marked by arrows. The absence of solar neutrino noise makes antineutrino detection an  ideal tool to disentangle the solar flare signal} \label{Fig2}
\end{center}
\end{figure}
\section{Why it is so relevant for us}

One of the basic reason to have a neutrino solar flare astronomy is to discover in an independent way the
solar neutrino flavor oscillation and mix, even taking into account the MSW effect in peculiar configuration.
However, the social argument in favor of a solar neutrino detection is the possibility to have a prompt alarm system
both for any robotic, in orbit satellite (for communication, military, scientific detection) but in particular for the
human pilot or astronaut in flight. The Earth magnetic field may partially screen such a huge solar flare, but an astronaut in flight, for instance, just to our Moon are at life risk. The possibility to have a very rapid  alarm may allow the astronauts, for example, to immerse themselves immediately inside an inner large water reserve container, able to screen them from the intense GeVs energy solar wind that would strike half an hour later. The Martial mission may deeply depend on the ability to foresee, inform and protect the astronauts flight by a fast and prompt solar neutrino detection \cite{Fargion_2011}.

\begin{figure}[!t]
\begin{center}
\includegraphics[scale=0.49]{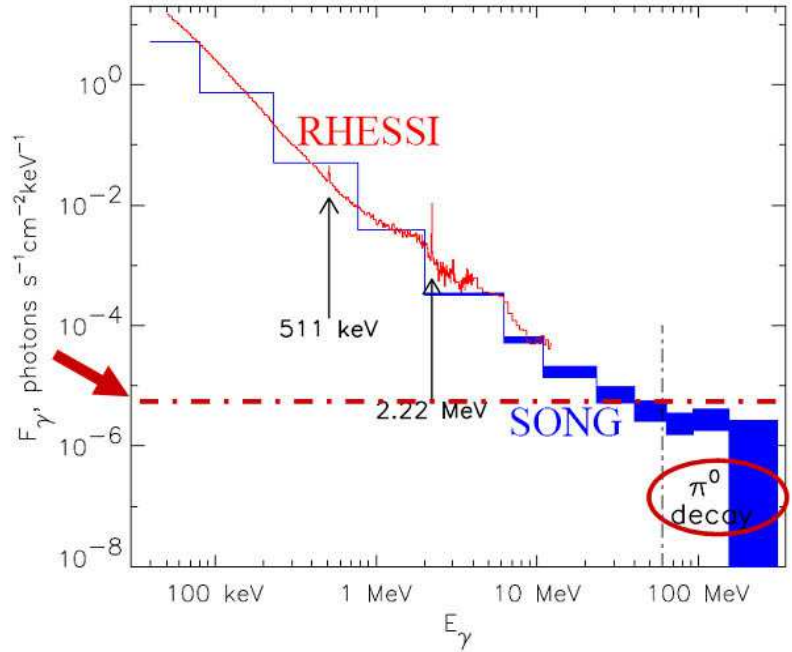}
\caption{The observed gamma spectra in brightest solar flare offering a lower bound in neutral pion production and consequently the lower bound for solar flare neutrino} \label{Fig3}
\end{center}
\end{figure}
\section{Where and why solar neutrino flare may be soon observed}

Indeed large size neutrino detectors (1, 2, 10 Megaton) at GeVs up to ten GeV (HK, PINGU, DeepCore) are just
recording data or are under construction; ten GeV is a moderate high energy but hardest solar
flare have been observed up to tens GeV gamma energy. Moreover, even a few GeV  detectors, may soon be born (HK japan, HK Korea). Our recent estimate based on a lower (gamma) bound of neutrino signal, while below the upper ones, still confirms the order
of magnitude and the near edge discovery. The recent peculiar solar flares as the October-November
2003 and January 2005  \cite{Fargion_2006}  were source of high energetic charged particles: large fraction of these primary particles,
became a source of both neutrons \cite{Fargion_2004} and secondary kaons, $K$, pions, $\pi^{\pm}$ by their particle-particle spallation
on the Sun surface. Consequently,  final secondaries muon and electronic neutrinos and anti-neutrinos, are released by the chain reactions
occurring on the sun atmosphere. There are two different sites for these decays: 1) A brief and sharp solar
flare, originated within the solar corona itself and 2) a diluted and delayed terrestrial neutrino flux, produced by late flare particles hitting the Earth's atmosphere. This latter
delayed signal is of poor physical interest, like an inverse missing signal during the Forbush phase. The
main and first solar flare neutrinos reach the Earth with a well defined directionality and within a narrow time range.
Their corresponding average energies  suffer negligible energy loss. The opposite occur to downward flare. In the
simplest approach, the main source of pion production is the common scattering
$p+p$ making a $\Delta^+$ mostly at its center of mass (of the resonance) whose mass value is
$m_{\Delta^+}=1232$~MeV. As a first approximation and as a useful simplification after the needed boost of the secondaries
energies one may assume that the total pion Delta energy is equally distributed, in average, in all its four final remnants each at an average 30~MeV energy. The consequent spectra for largest solar flare  are described with the corresponding future detector mass (1, 5, 10 Megaton) threshold  respectively for HK in Japan, HK in Korea, PINGU and DeepCore in IceCube, see Fig.\ref{Fig1}.

\subsection{Solar neutrino flares above GeV energy: the muon track amplifier}

The near future and the present Megaton neutrino detector might be able to reveal the
solar neutrino flare; as shown in Fig.\ref{Fig1}. one may note that the imprint of the anti neutrino electron
 is much clean because of the silent solar thermonuclear emission or noise Fig.\ref{Fig1b}. This imply that also Gadolinium enhanced Megaton detector may play a role in Solar Flare and possibly in Relic SN neutrinos signals.  Above the GeV energy the appearance of the first solar flare muon neutrinos will be revolutionary: it has never been observed the second lepton signal, the muon neutrino, by any star, yet. Moreover there is a remarkable enhanced effect for muons.  The muon track in water extend nearly 5 meters for each GeV of energy, almost in linear growth. The possibility to observe along the edges of the SK or HK container external muon (through going muons) that are pointing to the sun during the flare may increase the effective volume by a significant factor. For instance at 10 GeV the eventual solar flare may double the SK volume and (because of the 2.5 rock density), it may even make three times a larger detection volume. The same amplify effect in HK is less remarkable because of the wider HK container sizes; nevertheless at several GeV muon tracks may amplify solar neutrino flare detection even twice the nominal detector mass. See Figg.\ref{Fig2}, \ref{Fig3}; in comparison note the observed  large gamma flare in Fig.\ref{Fig3}

 \section{Conclusions}
The solar flare neutrino is at hand. Its detection is well within Deep Core, Pingu, HK, detector system.
The absence of detection in SK may be due to a low mass density in the outer flare propagation mass ejection.
However there is also a prompt gamma signal that imply a neutrino one that it may or better say must be observed in largest Megaton neutrino detector. Its discover will offer the first nearest neutrino hadronic imprint in modern neutrino Astronomy.



\begin{thebibliography}{99}
\bibitem{Fargion_2003}  Fargion,D., Moscato, F. Chin.J.Astron.Astrophys.3,S75-S76.(2003)
\bibitem{Fargion_2004}  Fargion,D, JHEP06,045;(2004)
\bibitem{Fargion_2006}  Fargion, D., Phys.Scripta,T127,(2006),22-24
\bibitem{Grechnev 2008} Grechnev V.V. et al., Solar, Physics,Vol.1,October,252,(2008)
\bibitem{Fargion_2011} Fargion, D.; ICRC32, Vol. 6, OG1.4, 303, 2011











%
%
%
%
%
%
%
%
%
%
%
%
%
%
%


%
%
%
%
%
%
%
\end{thebibliography}
\end{document}